\documentclass[aps,prb,twocolumn,showpacs,superscriptaddress]{revtex4}
\usepackage{graphicx}
\usepackage{amsfonts,amssymb,amsmath}
\usepackage{bm}
\begin{document}

\title{A test of the spin-orbit sum rule for actinides using an {\it ab-initio}
calculation}

\date{\today}
 
\author{Matej Komelj}
\email{matej.komelj@ijs.si}
\affiliation{Jo\v zef Stefan Institute, Jamova 39, SI-1000 Ljubljana,
  Slovenia}
\begin{abstract}
The expectation value of the angular part of the spin-orbit-coupling (SOC) 
operator $\langle l\cdot s\rangle$ 
and the branching ratio 
for the $N_{4,5}$ X-ray-absorption spectroscopy in $\delta$-Pu, $\alpha$-U
and  face-centered-cubic (fcc) Th are calculated within the framework of the 
density-functional theory in  combination with the 
local-spin-density approximation (LSDA), the generalized-gradient approximation 
(GGA) and the LDA+$U$ method.  The SOC contribution is calculated in terms
of the first- and the second- variational schemes.
A strong variation in the magnitudes
of the calculated $\langle l\cdot s\rangle$ operator along the actinide 
series is found.
The results show that the SOC sum rule for 
$\delta$-Pu and $\alpha$-U is reasonably valid, whereas there is a sign 
mismatch for the case of fcc 
Th. The calculated branching ratios for the case without SOC in the valence 
shell strongly deviate from the statistical value. 
\end{abstract}
\maketitle
The field of actinides attracts considerable interest in terms of 
experiments and theoretical investigations.
Recently, many unique phenomena of the actinide
series were discovered, for example, the Pu-based superconductivity 
\cite{Sarrao:2002,Opahle:2003,Maehira:2003} or the phonon dispersion in 
$\delta$-Pu \cite{Lander:2003,Dai:2003,Wong:2003}, which were ascribed to the
nature of the $5f$ electronic states. In order to explain these
interesting properties, it is necessary to understand the electronic structure.
It was demonstrated 
experimentally that the spin-orbit coupling (SOC) could not be neglected 
in the Hamiltonian for the $5f$ states of 
the actinides\cite{Moore:2003}. \par
Van der Laan and Thole\cite{vanderLaan:1988} proposed  a probe for the 
SOC interaction, which is based on the branching ratio of the 
core-valence transition 
in X-ray absorption spectroscopy (XAS). The corresponding sum rule for the
$4d\rightarrow 5f$ transition, which is relevant for actinides, reads as:
\begin{equation}
{\langle l\cdot s\rangle\over N_h}=-{15\over 4}\left( B-B_0\right).
\label{sumrule}
\end{equation}
Here, $\langle l\cdot s\rangle$ denotes the angular part of the SOC operator
for a particular valence shell with $N_h$ holes.  The branching 
ratio $B$ for an electron transition from a $d$ core shell is defined as:
\begin{equation}
B={I_{5/2}\over I_{5/2}+I_{3/2}},
\end{equation}
where $I_{5/2}$ and $I_{3/2}$ are 
the total absorptions for the $d_{5/2}$ and $d_{3/2}$ levels, 
respectively. The quantity $B_0$ is the corresponding branching ratio if there
was no SOC in the valence shell. In the case of a negligible core-valence
interaction the statistical value of $B_0$ for a $d$ core shell is $3/5$. 
There are several problems related to the application of the sum rule 
(\ref{sumrule}) on experimental data. The number of holes $N_h$ for a particular
shell is not exactly known, while the quantity $B_0$ is not measurable, 
therefore it is necessary to rely on some estimates or hints from calculations.
Furthermore, the validity of Eq. (\ref{sumrule})  is
not guaranteed since it is based on a free-atom model. For example, it was
demonstrated\cite{Thole:1988} that
the SOC sum rule held well for the $4d$ and $5d$ metals but that it was
violated for the $3d$ and $4f$ metals due to the core-valence interactions.
The aim of the present paper is to test the validity of the SOC sum rule
for actinides by comparing the SOC expectation value
$\langle l\cdot s\rangle$ calculated directly from the electronic structure 
with the value obtained from the calculated branching
ratio via Eq. (\ref{sumrule}). 
A theoretical approach has several advantages over an experiment, because
such a test cannot be performed on measured data, and
both the experimentally uncertain quantities,
$N_h$ and $B_0$ can be easily determined with a calculation.\par
The calculations for $\delta$-Pu, $\alpha$-U and the face-centered-cubic (fcc) 
phase of Th were performed within the framework of the density functional
theory by applying the Wien97 code \cite{Blaha:1990}, which adopts the
full-potential linearized-augmented-plane-wave (FLAPW) method\cite{Wimmer:1981}.
The experimental values of the lattice parameters were used. The convergency
tests implied that 1000 ($\delta$-Pu and fcc Th) and 864 ($\alpha$-U) $k$ points in 
the full Brillouin zone (BZ) were enough for a calculation of the 
corresponding
quantities with the desired accuracy when the modified tetrahedron 
method\cite{Blochl:1994} was used for the BZ integration. The plane-wave
cut-off parameters were $6.7\>{\rm Ry}$ ($\delta$-Pu), $9.8\>{\rm Ry}$ 
($\alpha$-U) and $5.6\>{\rm Ry}$ (fcc Th).
\par
Since the 
correlation effects are important for 
actinides\cite{Savrasov:2001,Soderlind:2001,Bouchet:2000}, the influence 
of the local-spin-density approximation (LSDA)\cite{Perdew:1992}, 
the generalized-gradient approximation (GGA)\cite{Perdew:1992-2} and
the LDA+$U$ method\cite{Anisimov:93,Liechtenstein:95} was investigated.
The parameters $U$ and $J$, which appear in the LDA+$U$ scheme, were set to
$U=2\>{\rm eV}$ and $J=0.5\>{\rm eV}$ for all three systems. This universal 
choice was based on values found in the literature for plutonium (Ref. 
\onlinecite{Bouchet:2000}) and uranium (Ref. \onlinecite{Yaresko:2003}), and it 
was simply 
generalized to the case of thorium. 
The density-functional theory can yield a magnetic solution for 
actinides\cite{Stojic:2003}, hence the up and down  spin states were 
distinguished. 
However, for all three systems the
resulting magnetic moments were zero within the prescribed accuracy. 
The criterion for the self-consistency was the difference in the 
charge densities after the last two iterations being less than $10^{-4}\>
{\rm e/(a.u.)^3}$.\par
Special attention was paid to the calculation of the SOC contribution.
The standard approach within the FLAPW method is to apply the second 
variational scheme\cite{Singh:1994,Novak?}. In this method the eigenvalues
and the eigenvectors are calculated in two steps within each iteration. First,
a scalar relativistic Hamiltonian (without the SOC term) is diagonalized. 
Then, the lowest of the resulting scalar-relativistic eigenvalues and 
eigenvectors are used as a restricted basis set for diagonalizing the full
Hamiltonian with the included SOC term. 
While the second-variational method is sufficiently accurate,
for example, for the $3d$ transition metals,
it might break down for heavy elements like the actinides, as discussed in 
Refs. \onlinecite{Singh:1994,Nordstrom:2000}. A possible way to improve the 
results is to extend the basis of the second-variational step by including
relativistic $p_{1/2}$ local orbitals\cite{Kunes:2001}.
An alternative, more straight-forward, albeit less time-effective way, is 
to perform the calculations
in terms of the so-called first-variational scheme, where
the full Hamiltonian, including the SOC term, is diagonalized in a single
step, using the full basis set of the linearized-augmented plane waves.
In the present paper, the first- and the second-variational schemes were 
applied. 
The SOC term was simply set to 
zero for the calculation of the branching ratio $B_0$. \par
The XAS spectra 
$\mu_{5/2}(\epsilon)$ and $\mu_{3/2}(\epsilon)$ as a function of the photon 
energy $\epsilon$
were calculated using Fermi's golden rule in a nonrelativistic dipole 
approximation that is based on the evaluation of the matrix element
for the operator $\hat{\mathbf p}\cdot \mathbf{e}$ with
$\mathbf{e}$ denoting the polarization vector of the light. The total 
absorptions $n_{5/2}$ and $n_{3/2}$, which are required for the calculation
of the branching ratios $B$ and $B_0$ (2), were obtained from the integrals:
\begin{equation}
n_j=\int_{E_\text{F}}^{E_{\text{C}}}\mu_j(\epsilon)\>d\epsilon, 
\end{equation}
where $E_\text{F}$ and $E_\text{C}$ denote the Fermi energy and the upper edge
of the $5f$ valence band, respectively.  The latter quantity is defined
with the number of holes $N_h$ and the corresponding density of states:
$n_{5f}(\epsilon)$ as: 
\begin{equation}
N_h=\int_{E_\text{F}}^{E_\text{C}}n_{5f}(\epsilon)\>
d\epsilon. 
\end{equation}\par
The results are presented in Table 1. Common to all the considered systems is a
strong deviation of $B_0$ from the statistical value of $3/5$ ascribed to 
a non-negligible coupling between the core and valence electrons. 
\begin{table*}
\begin{ruledtabular}
\begin{tabular}{llrrrrrrr}
&&${\langle l\cdot s\rangle\over N_\text{h}}$&&$-{15\over 4}\left(B-B_0\right)$&&
$N_\text{h}$&&$B_0$\\ 
\hline
$\delta$-Pu&LSDA&-0.548&(-0.555)&-0.606&(-0.600)&8.785&(8.802)&0.698\\
&GGA&-0.547&(-0.554)&-0.605&(-0.599)&8.778&(8.796)&0.698\\
&LDA+$U$&-0.672&(-0.310)&-0.645&(-0.393)&8.729&(8.853)&0.699\\
$\alpha$-U&LSDA&-0.035&(-0.035)&-0.066&(-0.059)&11.430&(11.514)&0.701\\
&GGA&-0.034&(-0.035)&-0.068&(-0.066)&11.518&(11.496)&0.700\\
&LDA+$U$&-0.033&(-0.038)&-0.066&(-0.066)&11.537&(11.619)&0.699\\
fcc Th&LSDA&-0.002&(-0.002)&0.006&(0.008)&13.512&(13.524)&0.702\\
&GGA&-0.002&(-0.002)&0.005&(0.008)&13.516&(13.528)&0.702\\
&LDA+$U$&-0.001&(-0.001)&0.006&(0.009)&13.612&(13.622)&0.702
\end{tabular}
\end{ruledtabular}
\caption{A comparison between the directly calculated expectation 
value of the SOC operator $\langle w^{110}\rangle$ per 
number of holes $N_h$ in the $5f$ shell and the quantity $-{5\over 2}\left(
B-B_0\right)$, obtained from the calculated XAS spectra. The quantities 
$B$ and $B_0$  represent the branching ratios for
the cases with  and without SOC in the $5f$ valence shell, 
respectively.  The values in brackets were obtained by applying
the second-variational scheme, while the rest of the values resulted from 
the first-variational treatment of the SOC term.}
\end{table*}
The calculated $B_0$ is almost a constant for $\delta$-Pu, $\alpha$-U and fcc 
Th,
regardless of 
the method applied for the calculation of the exchange-correlation potential,
although there is a slight increasing trend from plutonium to thorium. 
This might be due to
the mixing of the $4d_{5/2}$ and $4d_{3/2}$ core states, because
the splitting  between the two core levels grows  along the actinide series,
as demonstrated in Fig. 1.
The magnitudes of the expectation values of the SOC operator 
$\langle l\cdot s\rangle$ obtained
from the calculated branching ratios $B$ and $B_0$ via Eq. (1) in general 
overestimate the corresponding quantities calculated directly from the 
electronic
structure. The agreement is the best in the case of $\delta$-Pu, within $\sim
4\%$ to $\sim 27\%$, depending on the calculational details.  While there
is almost no difference between the results obtained using the LSDA or GGA, and
the second-variational scheme yields almost identical results as the 
diagonalization of the full Hamiltonian with the included SOC term, the LDA+$U$ 
values differ considerably. The magnitudes of the $\langle l\cdot s\rangle$ 
operator
obtained by applying the first-variational method
are larger than the LSDA or GGA values. This is the
only case where the magnitude of the directly calculated 
$\langle l\cdot s\rangle$ operator is smaller
than the branching-ratio value, but the agreement is very good, within $4\%$. 
However, the situation is the opposite for the case where the second-variational
scheme was used for the calculation of the SOC term. The magnitudes are smaller 
than those obtained using the LSDA or GGA, and the magnitude of the 
branching-ratio SOC expectation value exceeds the directly calculated quantity 
by about 
$27\%$.  \par
The expectation values of the SOC operator for 
$\alpha$-U are one order of magnitude larger than the corresponding quantities 
for $\delta$-Pu, as it  would be expected on the basis of the XAS-spectra plot
in Fig. 1, where the difference between the $B$ and $B_0$ curves is 
substantially
less pronounced when compared to the case of $\delta$-Pu.
\begin{figure}
\includegraphics[width=.45\textwidth]{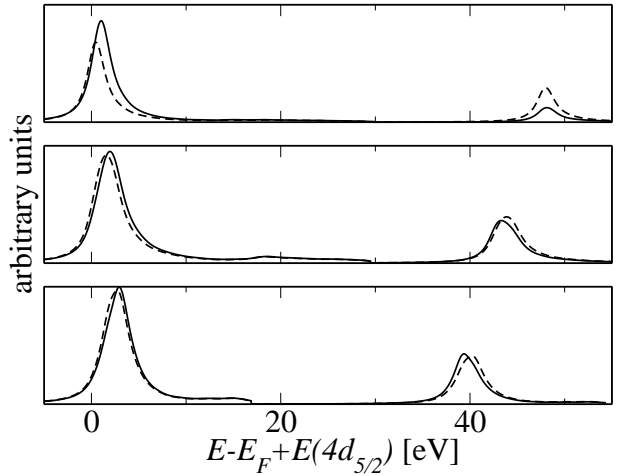}
\caption{The calculated total XAS spectra with (solid lines) and
without (dashed lines) SOC in the $5f$ valence shell for $\delta$-Pu 
(upper part), $\alpha$-U (middle part)
and fcc Th (lower part) obtained by applying the LSDA and the 
second-variational method.}
\end{figure}
The magnitudes of the branching-ratio values are about two times as large as 
those of the directly calculated quantities. The difference in the results 
obtained using the LSDA, GGA or the LDA+$U$ method is subtle, as is the 
influence of choice of the method for including the SOC term. \par
The expectation value of $\langle l\cdot s\rangle$ for fcc Th is reduced by 
another order of magnitude in comparison with $\delta$-Pu and $\alpha$-U. 
While the directly calculated quantity remains negative, the values obtained 
by applying Eq. (1) are positive, so that the SOC sum rule fails in this
case.  The reason is the absorption without SOC is larger than the absorption 
with SOC at the $N_4$ edge for fcc Th, in contrast to the case of $\delta$-Pu
or $\alpha$-U, as can be seen in Fig. 1.
If the statistical value of $3/5$ for $B_0$ was used instead of the 
result of the calculation, the sign of $\langle l\cdot s\rangle$ from the 
sum rule would be negative, as it is the directly calculated quantity, but
the order of magnitude would be wrong. The details of the calculation, 
namely the choice of the exchange-correlation potential and the SOC treatment,
do not have a significant influence on the results for fcc Th.\par 
In conclusion, the applicability of the SOC sum rule for actinides was 
investigated 
theoretically by comparing the  quantities obtained from the calculated
branching ratios with the quantities calculated directly from the electronic
structure. The agreement is the best for $\delta$-Pu, while the application
of the sum rule yields the wrong sign for fcc Th. It has to be noted that 
the corresponding quantities for $\delta$-Pu are two orders of magnitude
larger than those for fcc Th, hence the failure of the sum rule for the
latter case might be partly ascribed also to the limited accuracy of the 
calculation. It was found that neither the choice of the exchange-correlation
potential (the LSDA or GGA) nor the presence of the Coulomb repulsion 
(the LDA+$U$ method) had a substantial influence on the quantities under  
consideration.  The second-variational scheme was proven to be adequate for
the calculation of the SOC contribution for most of the 
considered cases. 
The exception is the LDA+$U$ calculation on $\delta$-Pu. The problem here is
with the mixed, spin-up and spin-down, matrix elements of the
orbital-dependent potential, which are supposedly too large to be taken into
account by means of the second-variational scheme. \par
The experimentalists should
bear in mind that the branching ratio $B_0$ for the case where the SOC 
interaction in the valence band is switched off strongly deviates from 
the statistical value. 
\bibliography{soc.bib}

\begin{thebibliography}{10}

\bibitem{Opahle:2003}
I. Opahle and P.~M. Oppeneer, Phys. Rev. Lett. {\bf 90},  157001  (2003).

\bibitem{Sarrao:2002}
J.~L. Sarrao, L.~A. Morales, J.~D. Thompson, B.~L. Scott, G.~R. Stewart, F.
  Wastin, J. Rebizant, P. Boulet, E. Colineau, and G.~H. Lander, Nature {\bf
  420},  297  (2002).

\bibitem{Maehira:2003}
T. Maehira, T. Hotta, K. Ueda, and A. Hasegawa, Phys. Rev. Lett {\bf 90},
  207007  (2003).

\bibitem{Lander:2003}
G.~H. Lander, Science {\bf 301},  1057  (2003).

\bibitem{Dai:2003}
X. Dai, S.~Y. Savrasov, G. Kotliar, A. Migliori, H. Ledbetter, and E. Abrahams,
  Science {\bf 300},  953  (2003).

\bibitem{Wong:2003}
J. Wong, M. Krisch, D.~L. Farber, F. Occelli, A.~J. Schwartz, T.~C. Chiang, M.
  Wall, C. Boro, and R. Xu, Science {\bf 301},  1078  (2003).

\bibitem{Moore:2003}
K.~T. Moore, M.~A. Wall, A.~J. Schwartz, B.~W. Chung, D.~K. Shuh, R.~K.
  Schulze, and J.~G. Tobin, Phys. Rev. Lett. {\bf 90},  196404  (2003).

\bibitem{vanderLaan:1988}
G. van~der Laan and B.~T. Thole, Phys. Rev. Lett {\bf 60},  1977  (1988).

\bibitem{Thole:1988}
B.~T. Thole and G. van~der Laan, Phys. Rev. B {\bf 38},  3158  (1988).

\bibitem{Blaha:1990}
P. Blaha, K. Schwarz, P. Sorantin, and S.~B. Trickey, Comput. Phys. Commun.
  {\bf 59},  399  (1990).

\bibitem{Wimmer:1981}
E. Wimmer, H. Krakauer, M. Weinert, and A.~J. Freeman, Phys. Rev. B {\bf 24},
  864  (1981).

\bibitem{Blochl:1994}
P.~E. Bl\"ochl, O. Jepsen, and O.~K. Andersen, Phys. Rev. B {\bf 49},  16223
  (1994).

\bibitem{Savrasov:2001}
S. Savrasov, G. Kotliar, and E. Abrahams, Nature {\bf 410},  793  (2001).

\bibitem{Soderlind:2001}
P. S\"oderlind, Europhys. Lett. {\bf 55},  525  (2001).

\bibitem{Bouchet:2000}
J. Bouchet, B. Siberchicot, F. Jollet, and A. Pasturel, J. Phsy.: Condens.
  Matter {\bf 12},  1723  (2000).

\bibitem{Perdew:1992}
J.~P. Perdew and Y. Wang, Phys. Rev. B {\bf 45},  13244  (1992).

\bibitem{Perdew:1992-2}
J.~P. Perdew, J.~A. Chevary, S.~H. Vosko, K.~A. Jackson, M.~R. Pederson, D.~J.
  Singh, and G. Fiolhais, Phys. Rev. B {\bf 46},  6671  (1992).

\bibitem{Anisimov:93}
V.~I. Anisimov, I.~V. Solovyev, M.~A. Korotin, M.~T. Czy\.zyk, and G.~A.
  Sawatzky, Phys. Rev. B {\bf 48},  16929  (1993).

\bibitem{Liechtenstein:95}
A.~I. Liechtenstein, V.~I. Anisimov, and J. Zaanen, Phys. Rev. B {\bf 52},
  R5467  (1995).

\bibitem{Yaresko:2003}
A.~N. Yaresko, V.~N. Antonov, and P. Fulde, Phys. Rev. B {\bf 67},  155103
  (2003).

\bibitem{Stojic:2003}
N. Stoji\'c, J.~W. Davenport, M. Komelj, and J. .Glimm, Phys. Rev. B {\bf 68},
  94407  (2003).

\bibitem{Singh:1994}
D.~J. Singh, {\em Plane Waves, Pseudopotentials and LAPW Method} (Kluwer
  Academic, Dordrecht, 1994).

\bibitem{Novak?}
P. Novak (unpublished).

\bibitem{Nordstrom:2000}
L. Nordstr\"om, J.~M. Wills, P.~H. Andersson, P. S\"oderlind, and O. Eriksson,
  Phys. Rev. B {\bf 63},  35103  (2000).

\bibitem{Kunes:2001}
J. Kune\v{s}, P. Nov\'ak, R. Schmid, P. Blaha, and K. Schwartz, Phys. Rev. B
  {\bf 64},  153102  (2001).

\end{thebibliography}
\end{document}